\ifx
\documentclass[conference]{IEEEtran}
\usepackage{amsthm}
\newtheorem{definition}{Definition}
\usepackage{graphicx}
\usepackage{mathtools,breqn}
\usepackage{tabularx,booktabs}
\usepackage{array}
\usepackage{dcolumn}
\usepackage{tabu}
\usepackage{algorithm}
\usepackage[noend]{algpseudocode}
\usepackage{amsfonts}
\usepackage{epsfig}
\usepackage{subfig}
\usepackage{subcaption}

\newcolumntype{F}{>{\bfseries}c}

\def\st{\textit{ s.t. }}

\fi

\documentclass[conference]{IEEEtran}
\usepackage{amsthm}

\usepackage{graphicx}
\usepackage{mathtools,breqn}
\usepackage{tabularx,booktabs}
\usepackage{array}
\usepackage{dcolumn}
\usepackage{subcaption}
\usepackage{multirow}
\usepackage{tabu}
\usepackage{algorithm}
\usepackage[noend]{algpseudocode}
\usepackage{amsfonts}
\usepackage{subfig}
\def\calc{\mathcal{C}}
\def\ovr{O}
\newcolumntype{F}{>{\bfseries}c}

\def\st{\textit{ s.t. }}

\ifCLASSINFOpdf
\else
\fi
\IEEEoverridecommandlockouts

\begin{document}
%\title{Toward Order-of-Magnitude Viral Cascade Prediction in Social Networks}
\title{Toward Order-of-Magnitude Cascade Prediction}
%Structural Diversity, Influence, and Diffusion in Social Networks}

\author{\IEEEauthorblockN{Ruocheng Guo \thanks{\noindent U.S. Provisional Patent 62/201,517. Contact shak@asu.edu for licensing information}, Elham Shaabani, Abhinav Bhatnagar and Paulo Shakarian}
\IEEEauthorblockA{%Ira A. Fulton Schools of Engineering\\
Arizona State University\\
Tempe, AZ\\
Email: \{rguosni, shaabani, abhatn, shak\}@asu.edu}}


\maketitle
\setlength{\textfloatsep}{5pt plus 1.0pt minus 2.0pt}
\begin{abstract}
When a piece of information (microblog, photograph, video, link, etc.) starts to spread in a social network, an important question arises: will it spread to ``viral'' proportions -- where ``viral'' is defined as an order-of-magnitude increase.  However, several previous studies have established that cascade size and frequency are related through a power-law - which leads to a severe imbalance in this classification problem.  In this paper, we devise a suite of measurements based on ``structural diversity'' -- the variety of social contexts (communities) in which individuals partaking in a given cascade engage.  We demonstrate these measures are able to distinguish viral from non-viral cascades, despite the severe imbalance of the data for this problem.  Further, we leverage these measurements as features in a classification approach, successfully predicting microblogs that grow from 50 to 500 reposts with precision of 0.69 and recall of 0.52 for the viral class - despite this class comprising under 2\% of samples.  This significantly outperforms our baseline approach as well as the current state-of-the-art. Our work also demonstrates how we can tradeoff between precision and recall.
\end{abstract}


% \fi
\IEEEpeerreviewmaketitle

\section{Introduction}

When a piece of information (microblog, photograph, video, link, etc.) starts to spread in a social network, an important question arises: will it spread to ``viral'' proportions -- where ``viral'' is defined as an order-of-magnitude increase.  Several previous studies~\cite{cheng2014can,shakBk} have established that cascade size and frequency are related through a power-law - which leads to a severe imbalance in this classification problem.  In this paper, we devise a suite of measurements based on ``structural diversity'' that are associated with the growth of a viral cascade in a social network.  Structural diversity refers to the variety of social contexts in which an individual engages and is typically instantiated (for social networks) as the number of distinct communities represented in an individual's local neighborhood.  Previously, Ugander et al. identified a correlation between structural diversity and influence~\cite{ugander2012structural}.  We demonstrate these measures are able to distinguish viral from non-viral cascades, despite the severe imbalance of the data for this problem.  Further, we leverage these measurements as features in a classification approach, successfully predicting microblogs that grow from 50 to 500 reposts with precision of 0.69 and recall of 0.52 for the viral class  (under 2\% of the samples).

We note that our results on the prediction of cascades rely solely upon the use of our structural diversity based measures for features and limited temporal features - hence the prediction is based on network topology alone (no content information was utilized).  We also achieved these results while maintaining the imbalances of the dataset - which we felt better mimics reality.  This differs from some previous studies which balance the data before conducting classification.  Further, we note that we obtained prediction of order-of-magnitude increases in the size of the cascade - which also differs from other work (i.e. \cite{cheng2014can}) which focus on identifying cascades that double in size.
\section{Technical Preliminaries}
\label{sec:prelim}

Here we introduce necessary notation and describe our social network data.  We represent a social network as a graph $G=(V,E)$ where $V$ is the set of vertices and $E$ as set of directed edges that have sizes $|V|,|E|$ respectively.  The intuition behind edge $(v,v')$ is that node $v$ can influence $v'$.  This intuition stems from how we create the edges in our network: $(v,v')$ is an edge if during a specified time period there is at least one microblog posted by $v$ that is reposted by $v'$ (we leave other thresholds beyond $1$ repost to future work).   We shall also assume a partition over nodes that specifies a community structure.  We shall assume that such a partition is static (based on the same time period from which the edges were derived) and that the partition $\calc$ consists of $k$ communities: $\{C_1, C_2, ...,C_k \}$.  There are many possible methods to derive the communities (if user-reported communities are not available).  We utilize the Louvain algorithm to identify our communities in this paper due to its ability to scale.

\noindent\textbf{Cascades.} For a given microblog $\theta$, we denote the subset of first-$m$ nodes who originally posted or reposted $\theta$ as $V_{\theta}^m$ and refer to them as \textit{adopters} (at size $m$).  Likewise, the set of reposting relationships within the same time period will be denoted $R_{\theta}^m$.  Taken together, we have a \textit{cascade}: $D_{\theta}^m = (V_{\theta}^m,R_{\theta}^m)$.  Any valid original microblog $\theta$ could be treated as a unique identifier for a cascade. Given a microblog $ \theta $, $v_{\theta}$ is the originator at instance $t_{\theta}^0$, which is defined as the origin time when the originator posted the microblog $\theta$ and time $t$ is time since $t_{\theta}^0$. The $m^{th}$ repost of the microblog $\theta$ happens at time $t_{\theta}^m$. As $m$ increases, a cascade accumulates nodes and edges over time.  We shall use $N_{\theta}$ to denote the final size of a cascade while the size of a cascade at any particular instance is the set of nodes present at that instance is simply $|V_{\theta}^m|$.  For a given size $m$, we shall refer to the \textit{frontiers} as the outgoing neighbors of the adopters in graph $G$ who are not adopters themselves.  Formally: $F_{\theta}^m = \left\{v \in V/V_{\theta}^m \st \exists v_i \in V_{\theta}^m \textit{ where } (v_i ,v) \in E  \right\}$.  For nodes in $G$ that are outside the adopters, we shall use the notation $t_{exp}(v,\theta,m)$ to denote the number of time units from the initial post of $\theta$ before the microblog was reposted by one of $v$'s incoming neighbors - intuitively the time at which $v$ was exposed to $\theta$.  For a given natural number $\lambda$ (used to specify a time period), we define the $ \lambda $ frontiers as a subset of the frontiers that have been exposed to $\theta$ no earlier than $\lambda$ time units previously.  Formally this set is defined as follows: $F_{\theta}^{m,\lambda} = \left\{v\in F_{\theta}^m | t_{exp}(v,\theta,m) \le \lambda \right\}$.  Finally, the complement of this set are the \textit{$\lambda $ non-adopters}: $\bar{F}_{\theta}^{m,\lambda} = \left\{v\in F_{\theta}^m | t_{exp}(v,\theta,m) > \lambda \right\}$.  

\noindent\textbf{Sina Weibo Dataset.}  The dataset we used was provided by WISE 2012 Challenge\footnote{http://www.wise2012.cs.ucy.ac.cy/challenge.html}. It included a sample of microblogs posted on Sina Weibo from 2009 to 2012. In this dataset, we are provided with time and user information for each post and subsequent repost which enabled us to derive a corpus of cascades.  From this data, we derived our social network $G=(V,E)$ (with 17.9 M vertices and 52.4 M edges) that was created from reposts that were published during the 3 month period between May 1, 2011 and July 31, 2011.
For this network, the average clustering coefficient is 0.107. There are 4974 connected components in the network.
Louvain algorithm outputs 379,416 communities with average size of 47.5 for this network.  
As expected, this network exhibits a power-law degree distrubtion.
For this network, the number of active nodes in August (the time period we studied for cascade prediction) is 5,910,608, while 5,664,625 of them at least have one out-neighbor.  During the month of August, there were 9,323,294 reposts with 2,252,368 different original microblogs. 1,920,763 ($86.6\%$) of them were written by authors who at least published one microblog during May 1, 2011 to July 31, 2011 (the time period we used to create the underlying network).
The average time it took for viral cascades to become viral is approximately 18 hours.
   The distribution of final size of cascades mimics a power-law distribution which can demonstrate that this dataset is more representative of cascade behavior observed ``in the wild''.  This differs significantly from the previous works which conduct biased sampling to artificially provide balanced classes.  We selected $\lambda$ as 30 minutes as 90\% of all reposts in the initial 3 month period occurred in under this time.

\noindent\textbf{Number of communities.} For $V' \subseteq V$, the associated communities $\calc(V')$ are the communities represented by $V'$.   Formally:  $\calc(V')=\{C_i \in \calc \st V' \cap C_i \neq \emptyset\}$.  The cardinality of this set (number of communities) will be denoted $K(V')$.  We measure the number of communities represented by the above three populations of nodes: $K(V_{\theta}^m)$, $ K(F_{\theta}^{m,\lambda})$, $K(\bar{F}_{\theta}^{m,\lambda})$ observed at either a given cascade size.

\noindent\textbf{Gini impurity.}  For $V' \subseteq V$, the gini impurity, $I_G(V')$ is the probability of a node in $V'$ being placed into the incorrect community if assigned a community based on the distribution of communities represented in $V'$.  Formally:
$I_G(V') = $\\$ 1 -\sum_{i}(\frac{\left\vert{C_i \cap V'}\right\vert}{\left\vert{V'}\right\vert})^2$.  We study the gini impurity of the adopters, $\lambda$ non-adopters and $\lambda$ frontiers for either a given cascade size $m$: $I_G(V_{\theta}^m)$, $I_G(F_{\theta}^{m,\lambda}) $, $I_G(\bar{F}_{\theta}^{m,\lambda})$.  The intuition is to capture a notion of how the communities are distributed amongst the nodes in each of these sets with a single scalar value.  We note that the impurity of the adopter set $I_G(V_{\theta}^m)$ behaves similar to the entropy of this set (a measurement introduced in \cite{weng2014predicting}).  However, as we will see in the next two sections, we found that the impurity of the $\lambda$ frontiers is a more discriminating feature.

\noindent\textbf{Overlap.} For $V_a, V_b \subset V$, the overlap ($\ovr(V_a,V_b)$) is simply the number of shared communities.  Formally: $\ovr(V_a,V_b)=|\calc(V_a)\cap \calc(V_b)|$.  We study overlap between adopters and $\lambda$ frontiers, between adopters and $\lambda$ non-adopters, and between $\lambda$ frontiers and $\lambda$ non-adopters: $O(V_{\theta}^{m},F_{\theta}^{m,\lambda})$, $O(V_{\theta}^{m},\bar{F}_{\theta}^{{m},\lambda})$, and $O({F}_{\theta}^{{m},\lambda},\bar{F}_{\theta}^{{m},\lambda})$ respectively.
The intuition with overlap stems directly from the original structural diversity results of \cite{ugander2012structural} - for instance a high overlap between adopters and $\lambda$ frontiers may indicate that the $\lambda$ frontiers are linked to adopters with inner-community connections and high structural diversity - hence increasing the probability of adoption.

\noindent\textbf{Average time to adoption.}  The average time to adoption for the nodes in the current set of adopters (once the cascade grows to size $m$): $ \frac{\sum_{i=1}^m t_{\theta}^i}{m} $. We also use average time to adoption as a baseline measure.

\section{Results}
\label{measSec}

Here we examine the behavior of the various structural diversity measurements as viral and non-viral cascades progress.  
We define a cascade as viral if the number of reposts reaches a threshold (denoted $TH$) of $500$ (in the next section we will explore other settings for $TH$ when describing our classification results).  
We look at snapshots of the cascades as they progress both in terms of size (denoted $m$).
For $m= \left\{10,30,50,100,200\right\}$, the number of samples is $\left\{98832,26733,13285,4722,1324\right\}$ respectively with 208 of the samples are viral.
With each size $ m $ we consider the Cascades with $ m $ adopters at some time $ t_{\theta}^m $, $ t_{\theta}^m $ can vary for different $ \theta $. Hence, cascades with final size $ N < m $ are ignored in our analysis task. This leads to a decrease in the number of non-viral Cascades as $ m $ increases.

\begin{figure}[!tbp]
	\begin{subfigure}[b]{0.23\textwidth}
		\includegraphics[width=\textwidth]{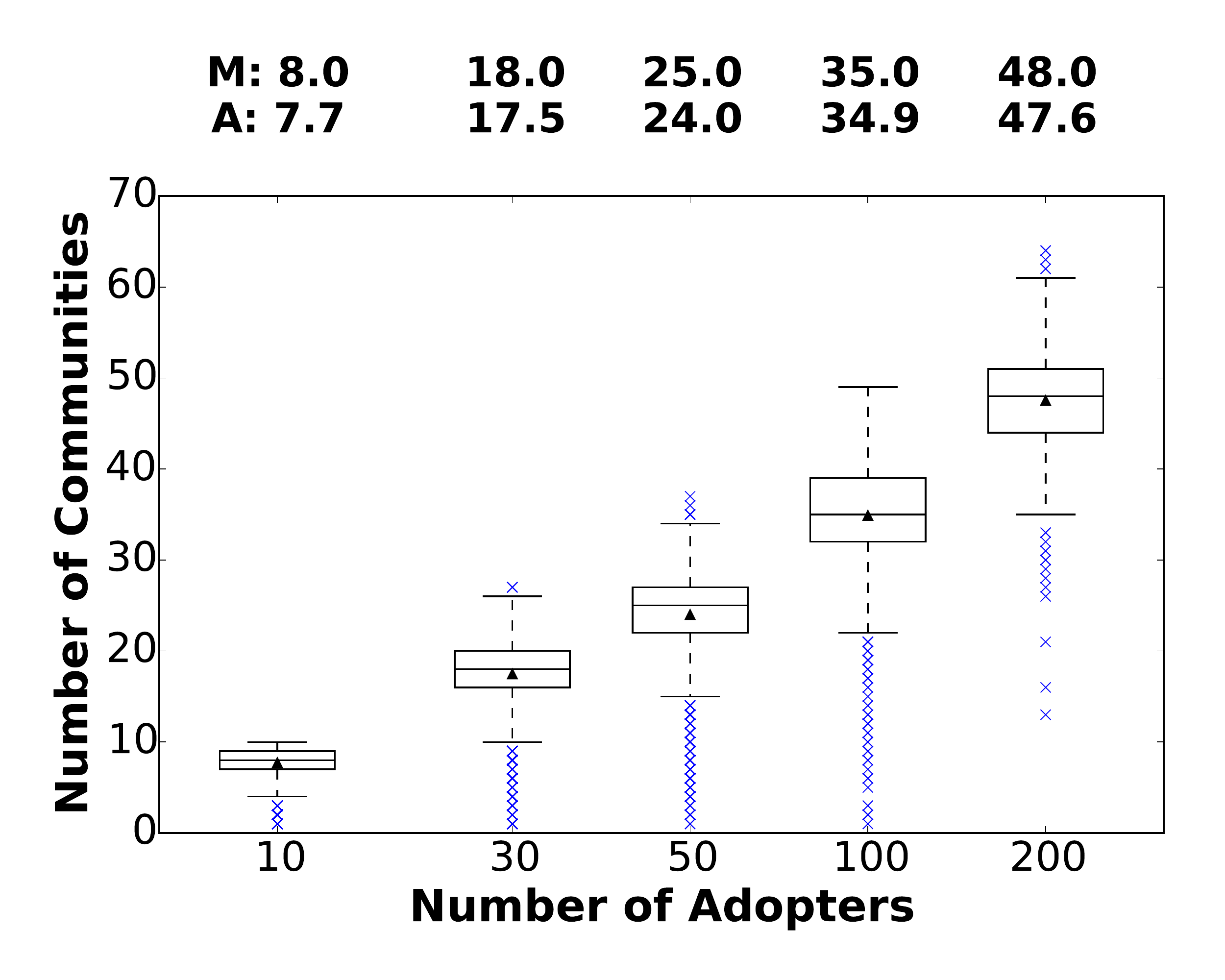}
		\caption{Number of communities amongst adopters ($K(V_{\theta}^{m})$) for non-viral cascades}
		\label{fig:S_nv_comm_a}
	\end{subfigure}
	\hfill
	\begin{subfigure}[b]{0.23\textwidth}
		\includegraphics[width=\textwidth]{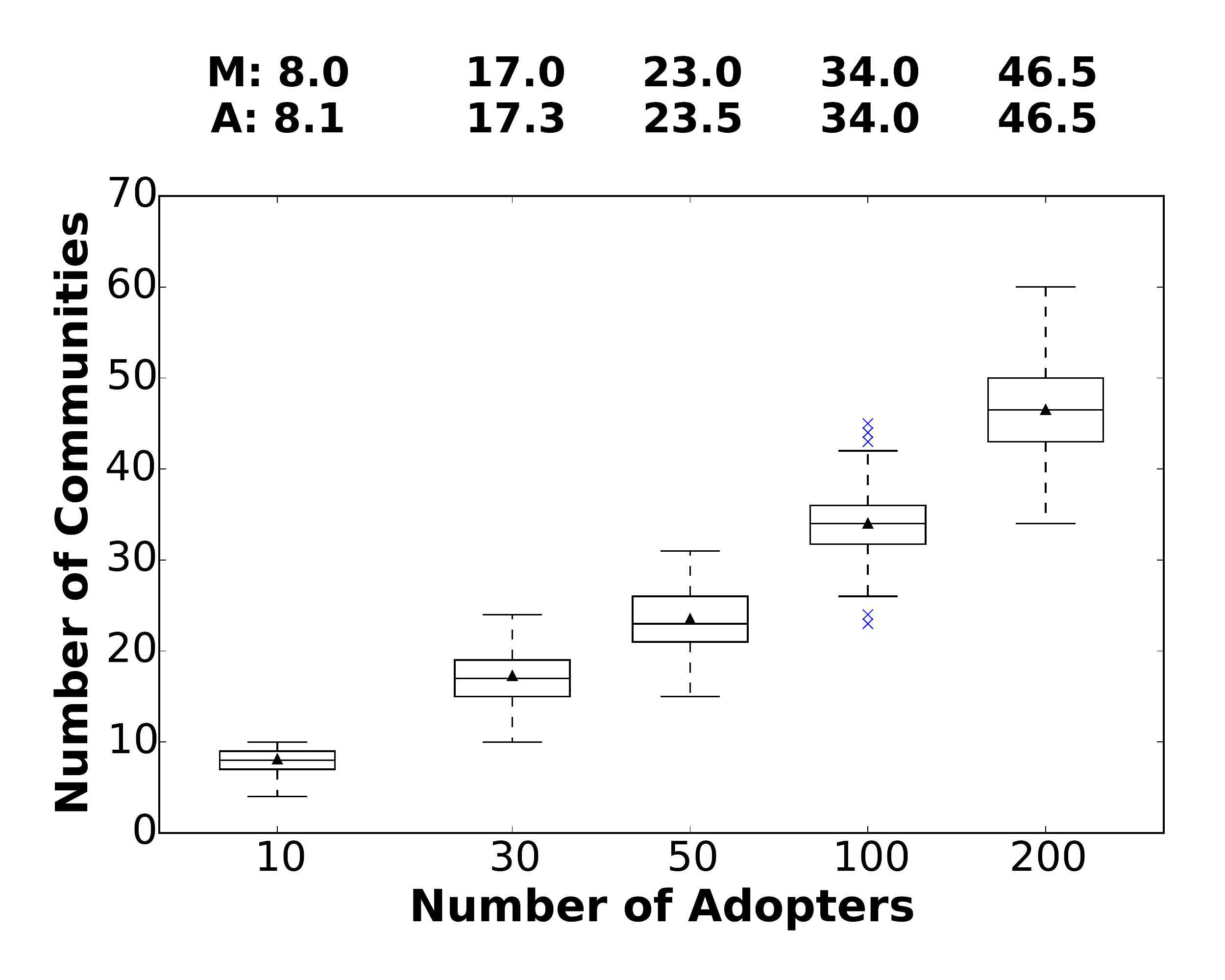}
		\caption{Number of communities amongst adopters ($K(V_{\theta}^{m})$) for viral cascades}
		\label{fig:S_v_comm_a}
	\end{subfigure}
	\hfill
	\begin{subfigure}[b]{0.23\textwidth}
		\includegraphics[width=\textwidth]{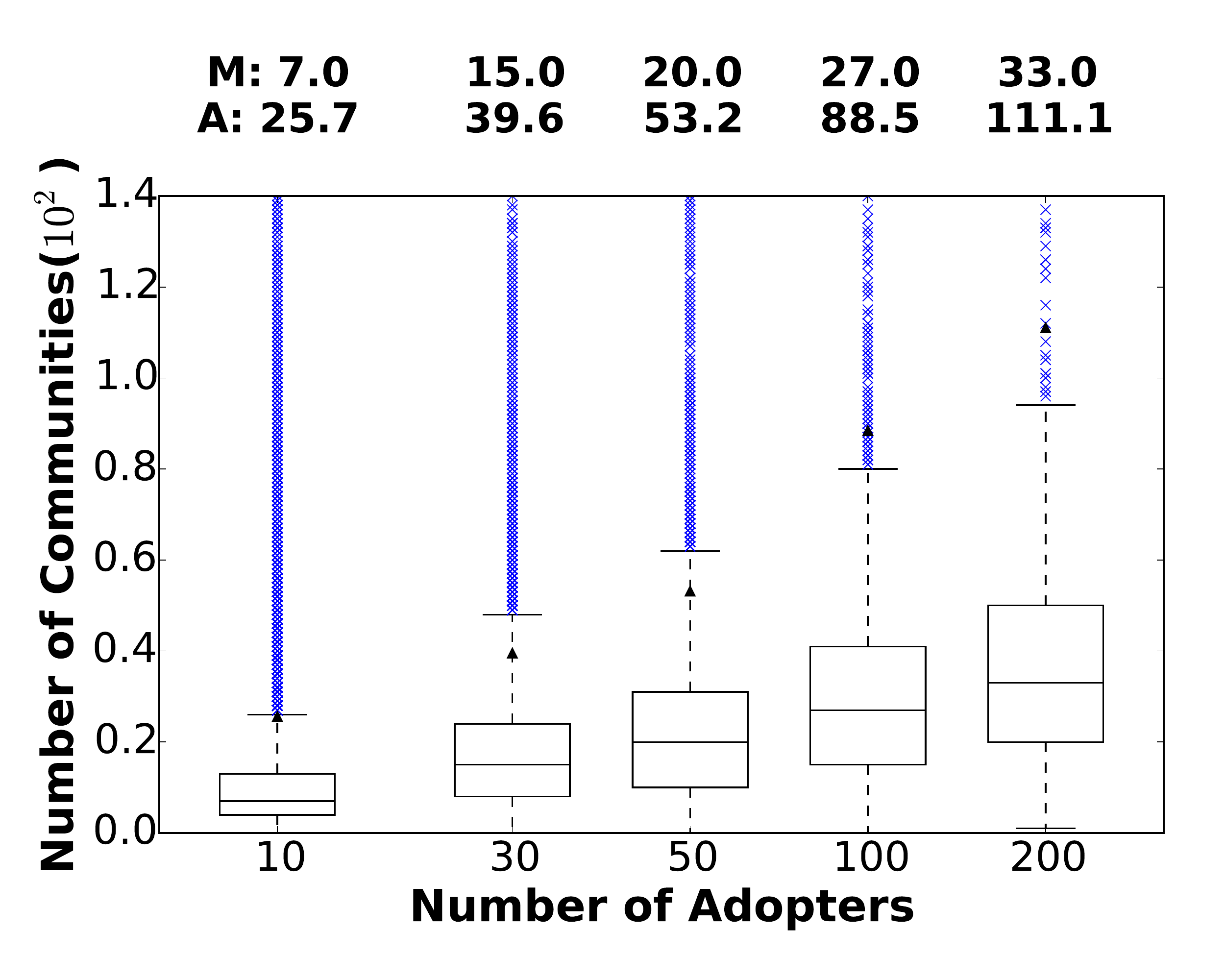}
		\caption{Number of communities amongst $\lambda$ frontiers ($K(F_{\theta}^{{m},\lambda})$) for non-viral cascades}
		\label{fig:S_nv_comm_f}
	\end{subfigure}
	\hfill
	\begin{subfigure}[b]{0.23\textwidth}
		\includegraphics[width=\textwidth]{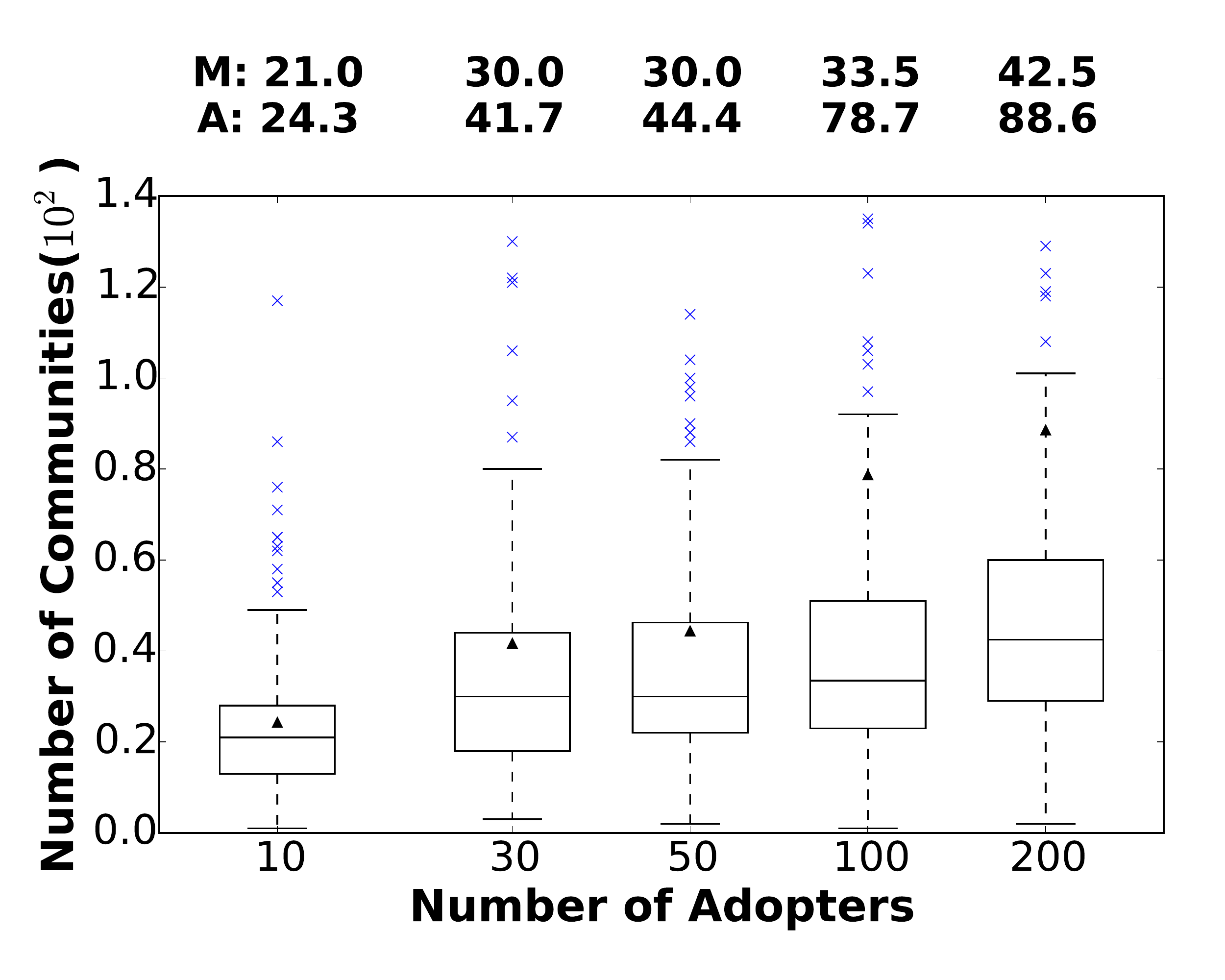}
		\caption{Number of communities amongst $\lambda$ frontiers ($K(F_{\theta}^{{m},\lambda})$) for viral cascades
		}
		\label{fig:S_v_comm_f}
	\end{subfigure}
	\begin{subfigure}[b]{0.23\textwidth}
		\includegraphics[width=\textwidth]{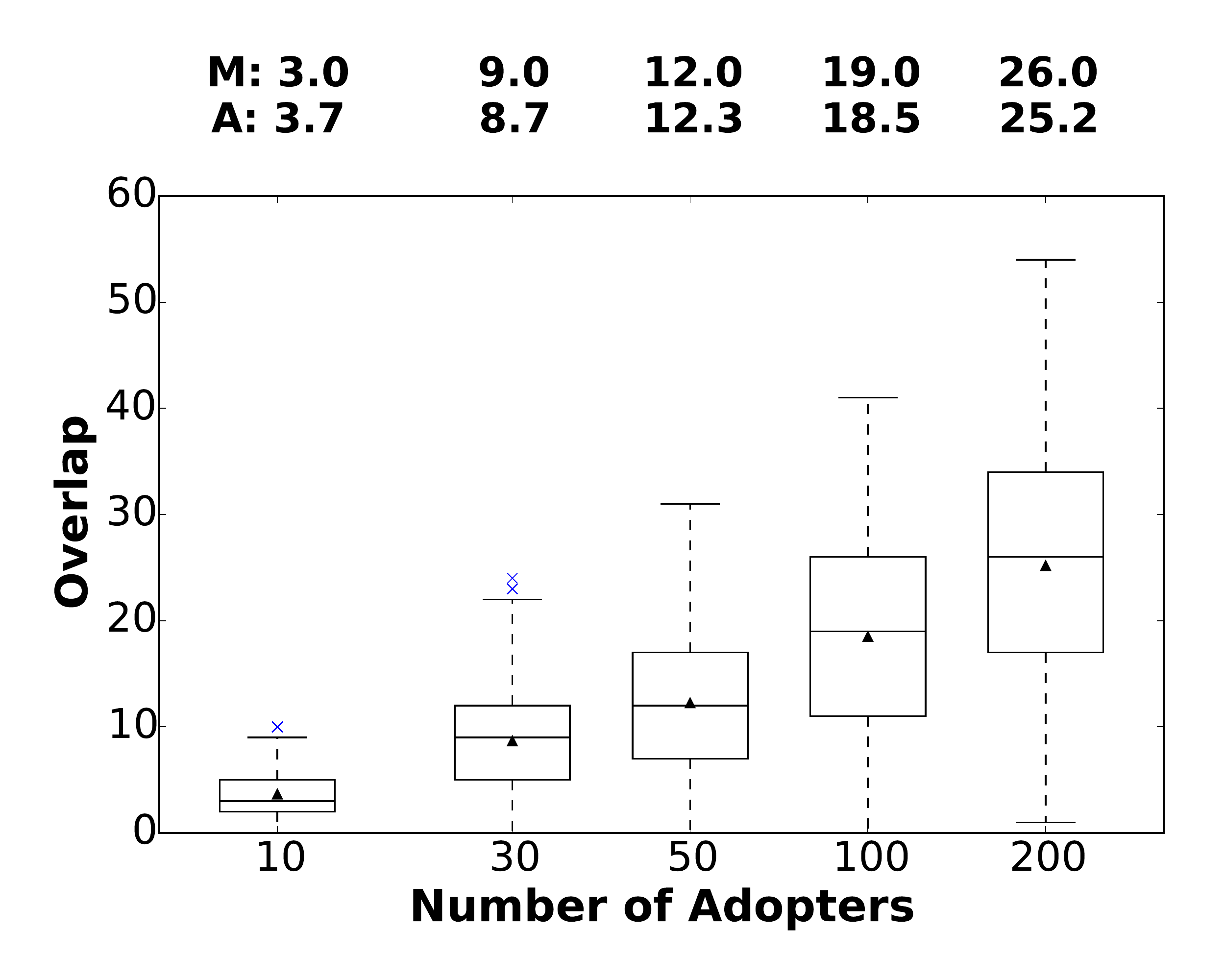}
		\caption{Overlap of adopters and $\lambda$ frontiers ($O(V_{\theta}^{m},F_{\theta}^{{m},\lambda})$) for non-viral cascades}
		\label{fig:S_nv_ol_af}
	\end{subfigure}
	\hfill
	\begin{subfigure}[b]{0.23\textwidth}
		\includegraphics[width=\textwidth]{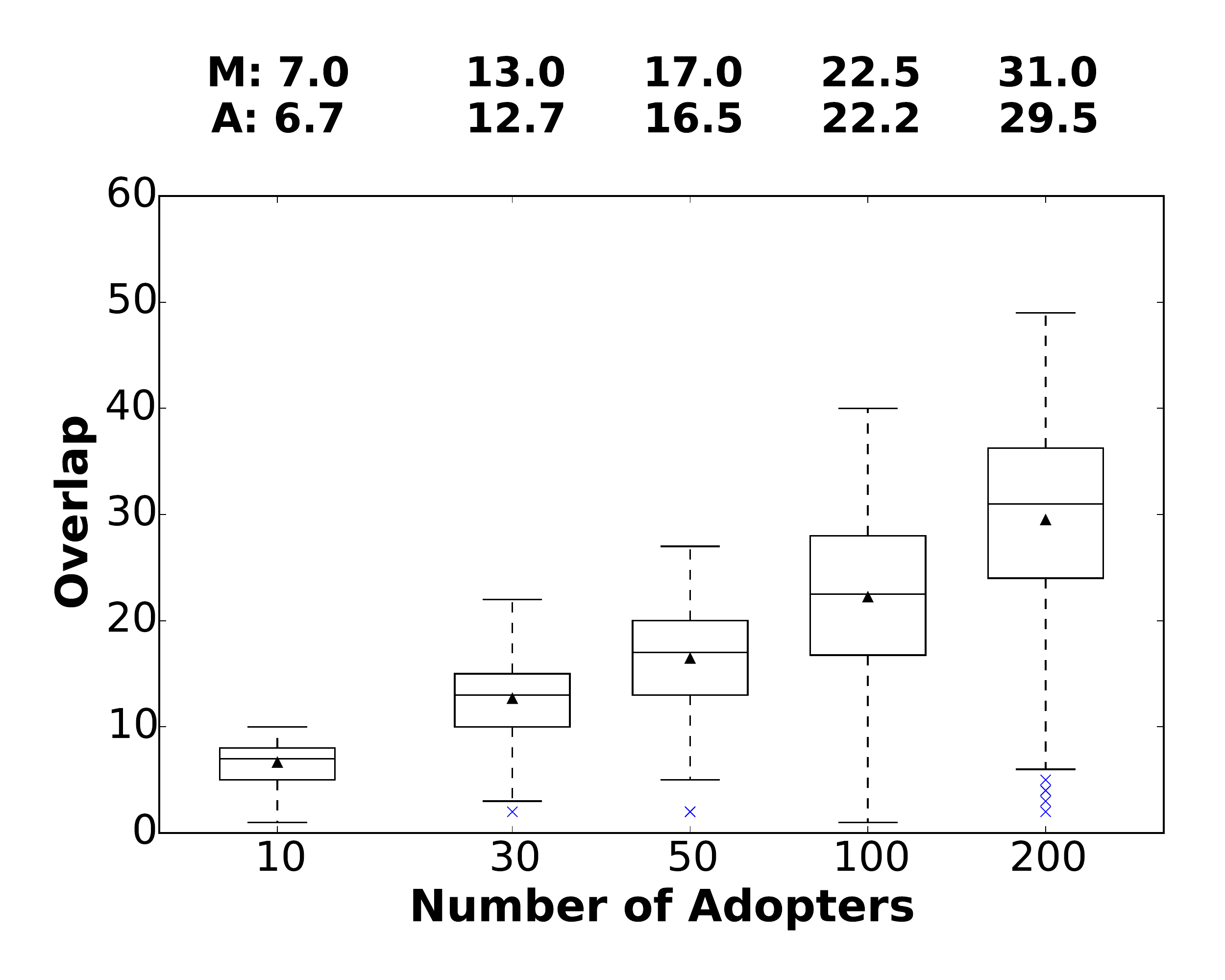}
		\caption{Overlap of adopters and $\lambda$ frontiers ($O(V_{\theta}^{m},F_{\theta}^{{m},\lambda})$) for viral cascades}
		\label{fig:S_v_ol_af}
	\end{subfigure}
	\begin{subfigure}[b]{0.23\textwidth}
		\includegraphics[width=\textwidth]{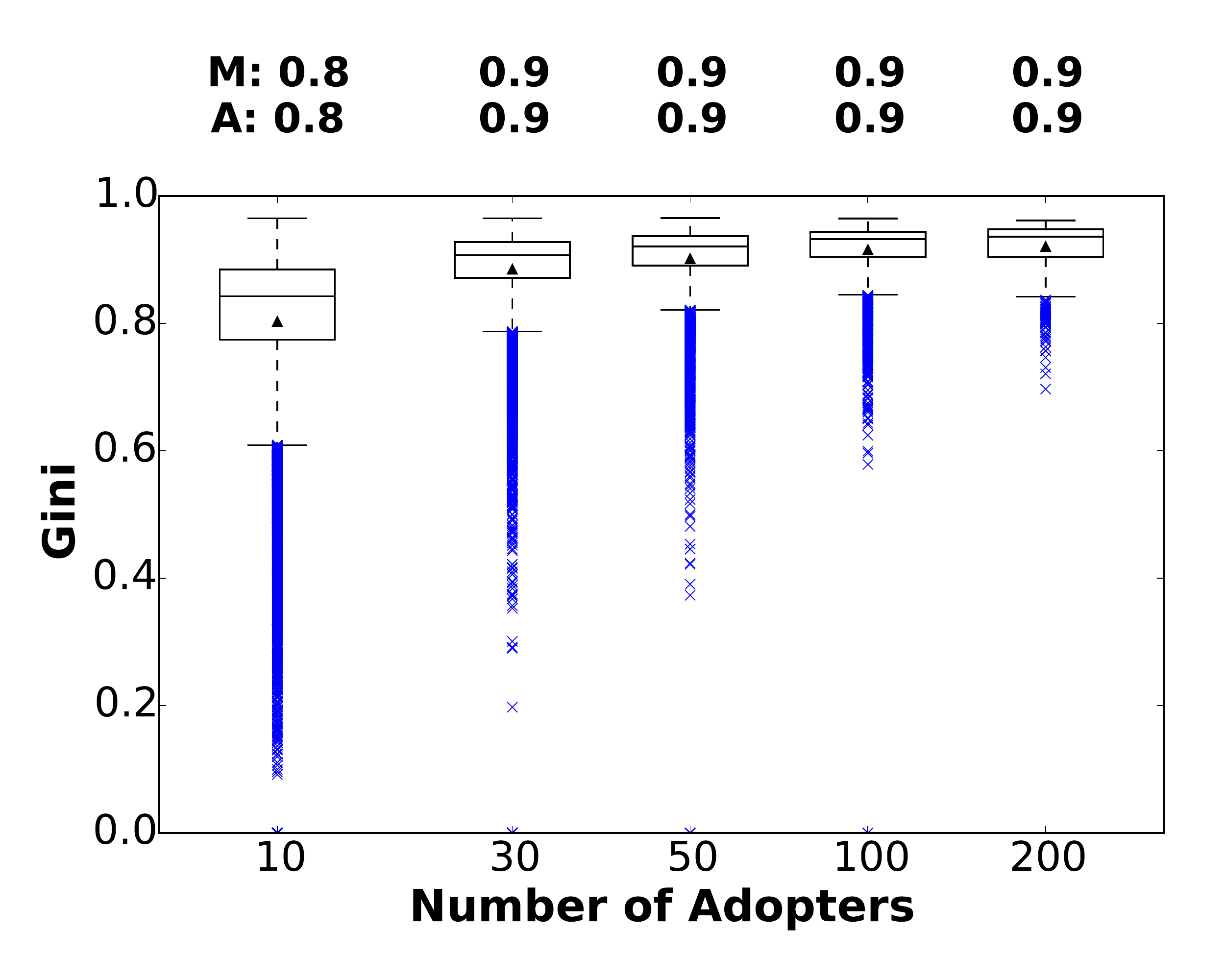}
		\caption{Gini impurity of $\lambda$ non-adopters ($I_G(\bar{F}_{\theta}^{{m},\lambda})$) for non-viral cascades}
		\label{fig:S_nv_ent_na}
	\end{subfigure}
	\hfill
	\begin{subfigure}[b]{0.23\textwidth}
		\includegraphics[width=\textwidth]{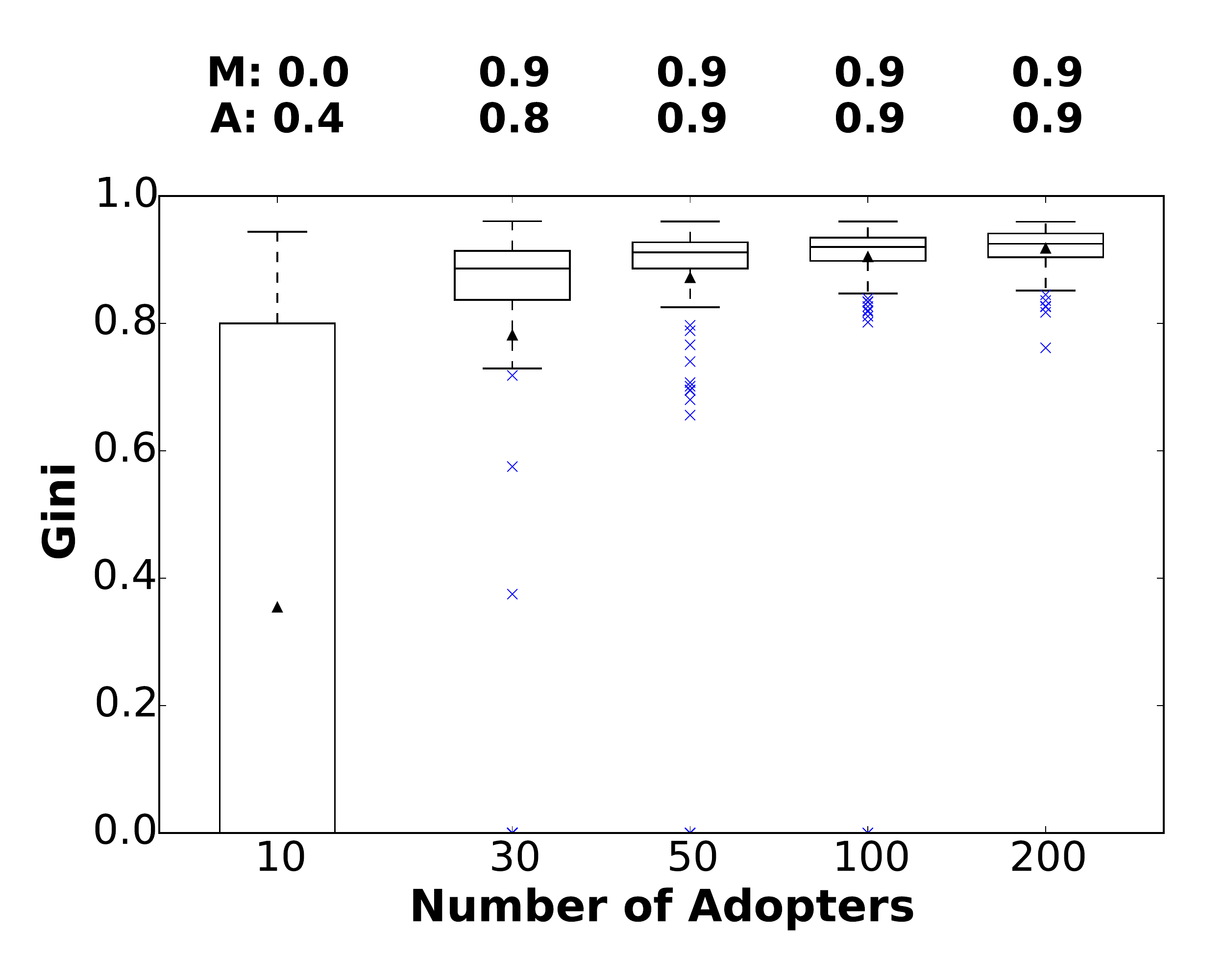}
		\caption{Gini impurity of  $\lambda$ non-adopters ($I_G(\bar{F}_{\theta}^{{m},\lambda})$) for viral cascades}
		\label{fig:S_v_ent_na}
	\end{subfigure}
	
	\begin{subfigure}[b]{0.23\textwidth}
		\includegraphics[width=\textwidth]{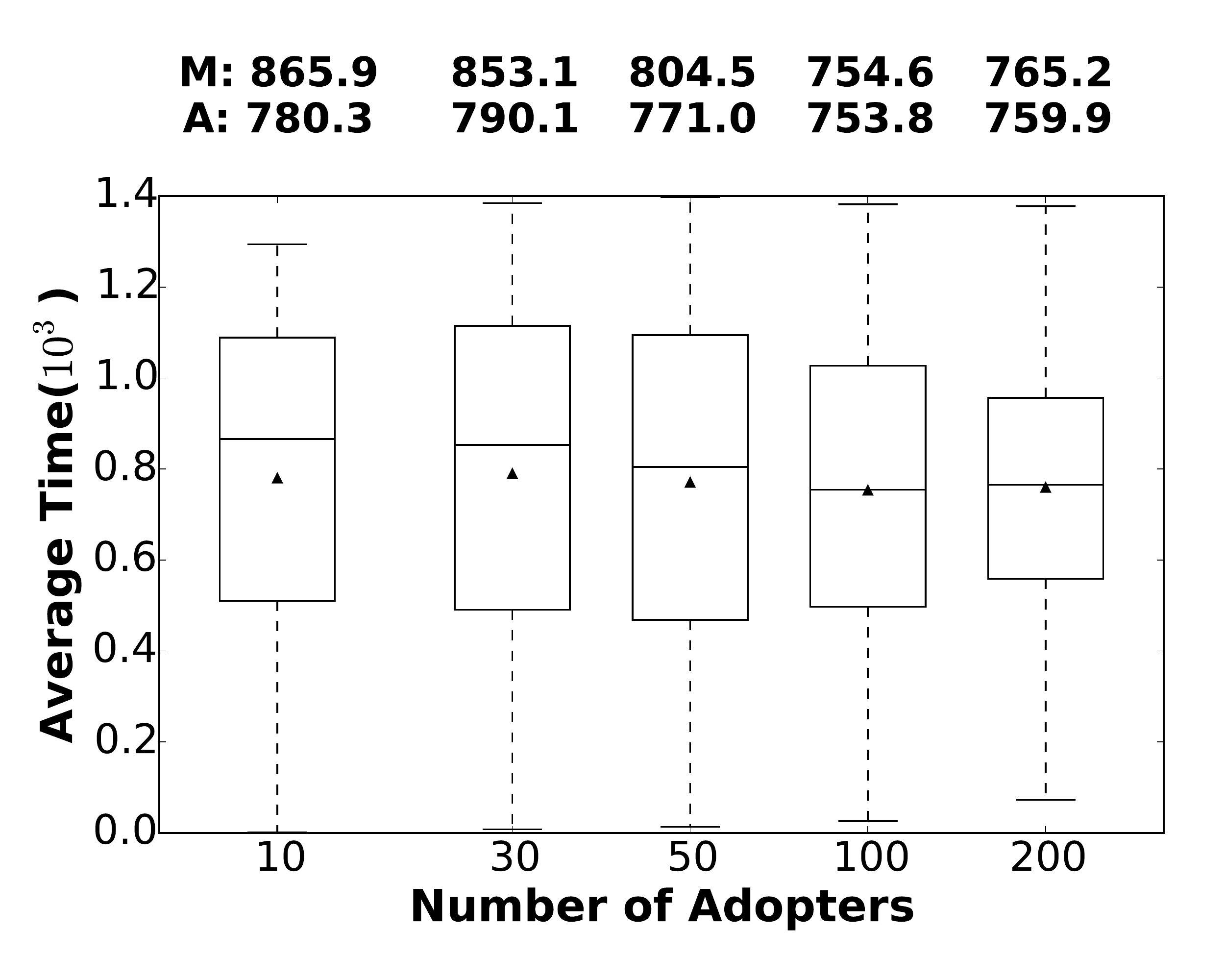}
		\caption{Non-viral cascades}
		\label{fig:nv_avg_time}
	\end{subfigure}
	\hfill
	\begin{subfigure}[b]{0.23\textwidth}
		\includegraphics[width=\textwidth]{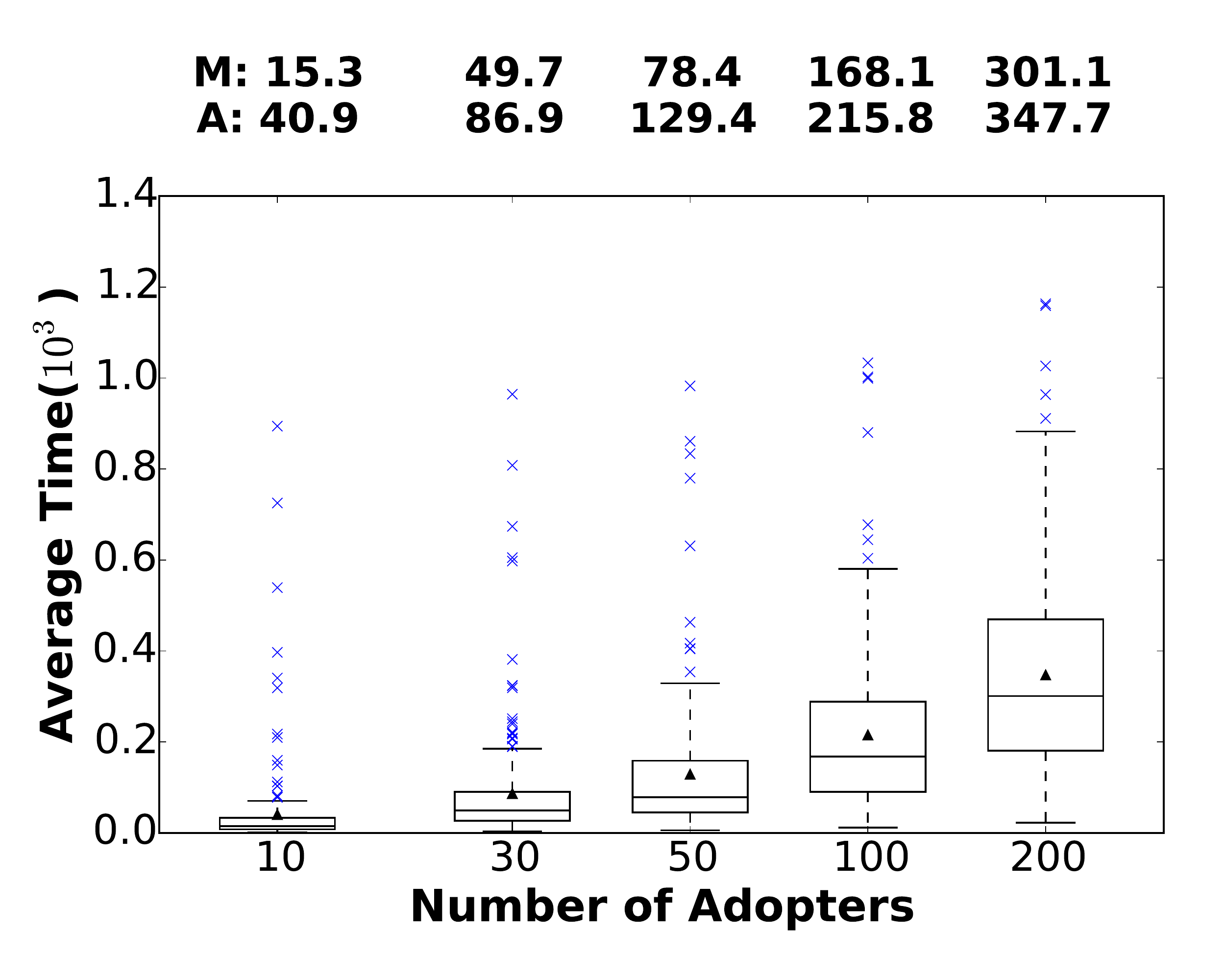}
		\caption{Viral cascades}
		\label{fig:v_avg_time}
	\end{subfigure}
	\caption{Number of communities, gini impurity, overlap and average time since $t_{\theta}^{0}$ to adoption for $ m = \left\{10,30,50,100,200\right\} $}
\end{figure}

\noindent\textbf{Average time to adoption.}  As a baseline measurement, we study the average time to adoption for each size-based stage of the cascade process (Fig.~\ref{fig:nv_avg_time}, Fig.~\ref{fig:v_avg_time}).  As expected, viral cascades exhibit a faster rate of reposting.  While we note that significant differences are present - especially in the early stages of the cascade, the whiskers of the non-viral class indicate a significant proportion of non-viral cascades that exhibit rapid adoption.  We believe this is likely due to the fact that certain cascades may have very high appeal to specialized communities.

\noindent\textbf{Number of communities.}  Fig.~\ref{fig:S_nv_comm_a}, Fig.~\ref{fig:S_v_comm_a}, Fig.~\ref{fig:S_nv_comm_f} and Fig.~\ref{fig:S_v_comm_f} display how the number of communities $ K(V') $ increases over $ m = \left\{10,30,50,100,200\right\} $ for the sets $ V' = \left\{V_{\theta}^m,F_{\theta}^{m,\lambda}\right\}$.  We note that $K(V_{\theta}^m)$ (the communities represented in the set of adopters) was shown to be a useful feature in \cite{weng2014predicting} for tasks where the target class had fewer reposts than in this study.  Here, we note that while statistically significant differences exist, the average and median values at each of the examined stages are generally similar.  On the other hand, the communities represented by the set of $\lambda$ frontiers ($K({F}_{\theta}^{{m},\lambda})$) shows viral Cascades have stronger capability than non-viral ones to keep a diverse set of $\lambda $ frontiers.  We also noted that the median of $K(\bar{F}_{\theta}^{{m},\lambda})$ (not pictured) shows viral cascades start with smaller $K({F}_{\theta}^{{m},\lambda})$. However, it increases faster in viral cascades as nodes in $\lambda $ frontiers becomes $\lambda $ non-adopters.

\noindent\textbf{Gini impurity.}  Cascades in both classes tend to accumulate diversity in the process of collecting more adopters - and we have also noted that a related entropy measure (studied in \cite{weng2014predicting}) performed similarly.  We also noted (not pictured) that in the early stages, viral cascades can show more diversity in $\lambda $ frontiers measured by $ I_G({F}_{\theta}^{{m},\lambda}) $ ($ m = \left\{10,30,50\right\} $).
But, perhaps most striking, that non-viral Cascades gain more uniformly distributed nodes over communities in $\lambda $ non-adopters, shown by $ I_G(\bar{F}_{\theta}^{{m},\lambda}) $ (Fig.~\ref{fig:S_nv_ent_na}, Fig.~\ref{fig:S_v_ent_na}).  We believe that this is due to non-viral cascades likely have an appeal limited to a relatively small number of communities - hence those \textit{not} adopting the trend may represent a more diverse set of communities.

\noindent\textbf{Overlap.}  We found that overlap grows with the number of adopters in the three types of overlap considered.  For $O(V_{\theta}^{m},F_{\theta}^{m})$, viral cascades start with a larger initial value and keep leading non-viral ones in the diffusion process of first 200 nodes (Fig.~\ref{fig:S_nv_ol_af}, Fig.~\ref{fig:S_v_ol_af}).  This may hint that viral cascades also take advantage of the densely linked communities to help them become viral.  However, in the case of $O(V_{\theta}^{m},\bar{F}_{\theta}^{m})$ and $O(F_{\theta}^{{m},\lambda},\bar{F}_{\theta}^{m,\lambda})$, viral cascades begin with lower value but grow much faster than non-viral Cascades.

\noindent\textbf{Classification Experiments.} 
Here we examine our experiments for predicting whether a cascade becomes viral - when a size threshold ($TH$) exceeds 500 adopters given that the cascade has 50 adopters ($ s = 50 $).  
Based on the distribution of final size of cascades in this dataset, this is a binary classification task with two heavily imbalanced classes.  
\textit{Hence, we report performance measurements (precision, recall and F1 score) for only the minority (viral) class.} Throughout the course of our experiments, we found that varying  threshold (slightly modifying the definition of ``viral'') for \textit{only} the training set allows for a trade-off between precision and recall.  
We study the trend of performance measures in two cases: 
(1.) The threshold for test set is maintained as $ TH_{ts} = 500 $ while the training threshold is varied $ TH_{tr} = \left\{300,400,500,600,700\right\} $.
(2.) The two thresholds are kept as the same $ TH $ while we modify this value $ TH = \left\{300,400,500,600,700\right\} $.

Table~\ref{tab:features} shows the groups of features used in our prediction tasks.  The features introduced in this paper is group $A_m$. As a baseline method for size-based prediction (feature group $C_m$) we used average time to adoption.  We also compare our features (Group $A_m$) with the community features extracted in \cite{weng2014predicting} (Group $B_m$).  This was the best performing feature set in that paper for a comparable task.\footnote{This was their highest-performing set of features for predicting cascades that grew from $50$ to $367$ and $100$ to $417$ reposts.  We also included the baseline feature in this set as we found it improved the effectiveness of this approach.}
Additionally,
we study the average size of recalled and non-recalled viral cascades by classifiers using features in groups $A_m$.
We also investigate the significance and performance of individual and certain combinations of features introduced in this paper.

\begin{table}[!tbp]%
	\setlength\extrarowheight{5pt}
	\renewcommand{\arraystretch}{1}
	\caption{\textmd{Features: Cascade Prediction over Time and Size}}
	\label{tab:features}
	\centering
	%\resizebox{\columnwidth}{!}{
	\begin{tabular}{| c| p{7cm}|}
		\hline
		\textbf{Group} & \textbf{Feature(s) over size } \\ \hline 
		\hline
		
		\multirow{3}{*}{$A_m$} & $ K(F_{\theta}^{m,\lambda})$,$K(\bar{F}_{\theta}^{m,\lambda})$,$I_G(V_{\theta}^{m})$,$I_G(F_{\theta}^{m,\lambda}) $,$I_G(\bar{F}_{\theta}^{m,\lambda})$,\\
		& $O(V_{\theta}^{m},F_{\theta}^{m,\lambda})$,$O(V_{\theta}^{m},\bar{F}_{\theta}^{m,\lambda})$,$O({F}_{\theta}^{m,\lambda},\bar{F}_{\theta}^{m,\lambda})$,\\
		
		& $|F_{\theta}^{m,\lambda}|$, $|\bar{F}_{\theta}^{m,\lambda}|$,
		$ \frac{\sum_{i=1}^m t_{\theta}^i}{m} $, $m \in \left\{30,50\right\}$
		\\ \hline
		$B_m$ & Community Features Mentioned in \cite{weng2014predicting} and $C_m$ \\ \hline 
		 \setlength\extrarowheight{5pt} $C_m$ & $ \frac{\sum_{i=1}^m t_{\theta}^i}{m} $, $m=50$ \\ \hline
	\end{tabular}
	%	}
\end{table}

We used ten-fold cross-validation in our experiments to ensure the results do not take any advantage of randomness in picking training and testing sets. First we carried out the prediction tasks with fixed thresholds $ TH_{tr} = 500$, $TH_{ts} = 500 $.
Then we modify the training threshold $ TH_{tr} =  \left\{300,400,500,600,700\right\}$ to show how this achieves a trade-off between precision and recall. The difference in average final size between correctly classified viral cascades and incorrectly classified ones is also monitored over $ TH_{tr} = \left\{300,400,500,600,700\right\}$ to show the potential to predict exact number of adopters by features.
Furthermore, we modify threshold of both training and testing sets $ TH = \left\{300,400,500,600,700\right\} $ to show the robustness of our features on related classification problems.
We used the oversampling method SMOTE with random forest classifier to generate synthetic samples for the viral class.  Other, lesser-performing classifiers were also examined (including SVM, MLP, and other ensemble methods) and are not reported here.  All results shown in this section is a sample mean produced by ten repeated experiments under each combination of variables.% Error bars represent one standard deviation.

\begin{figure}[!t]
	\centering
	\includegraphics[width=0.25\textwidth]{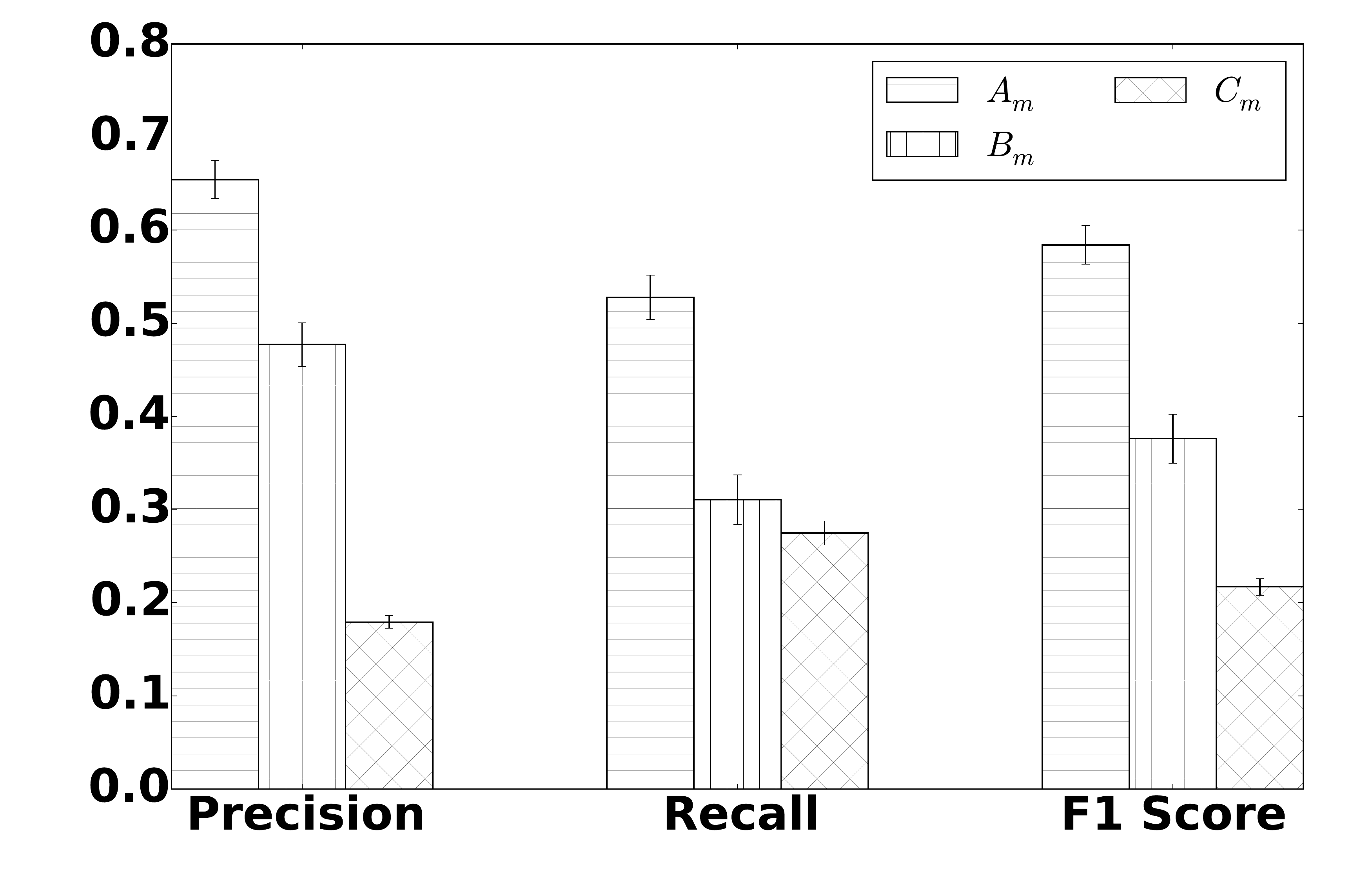}
	%\epsfig{figure=dgr,width=2.5in}
	%\label{fig:pred_SIZE}
	\caption{Classification results based on groups of features ($A_m$,$B_m$,$C_m$) extracted when $ m = 50 $ for fixed $ TH_{tr} = 500$, $TH_{ts} = 500 $. Error bars represent one standard deviation.}
	\label{fig:pred_Size}
	
\end{figure}

\begin{figure}[!tbp]
	\begin{subfigure}[b]{0.235\textwidth}
		\includegraphics[width=\textwidth]{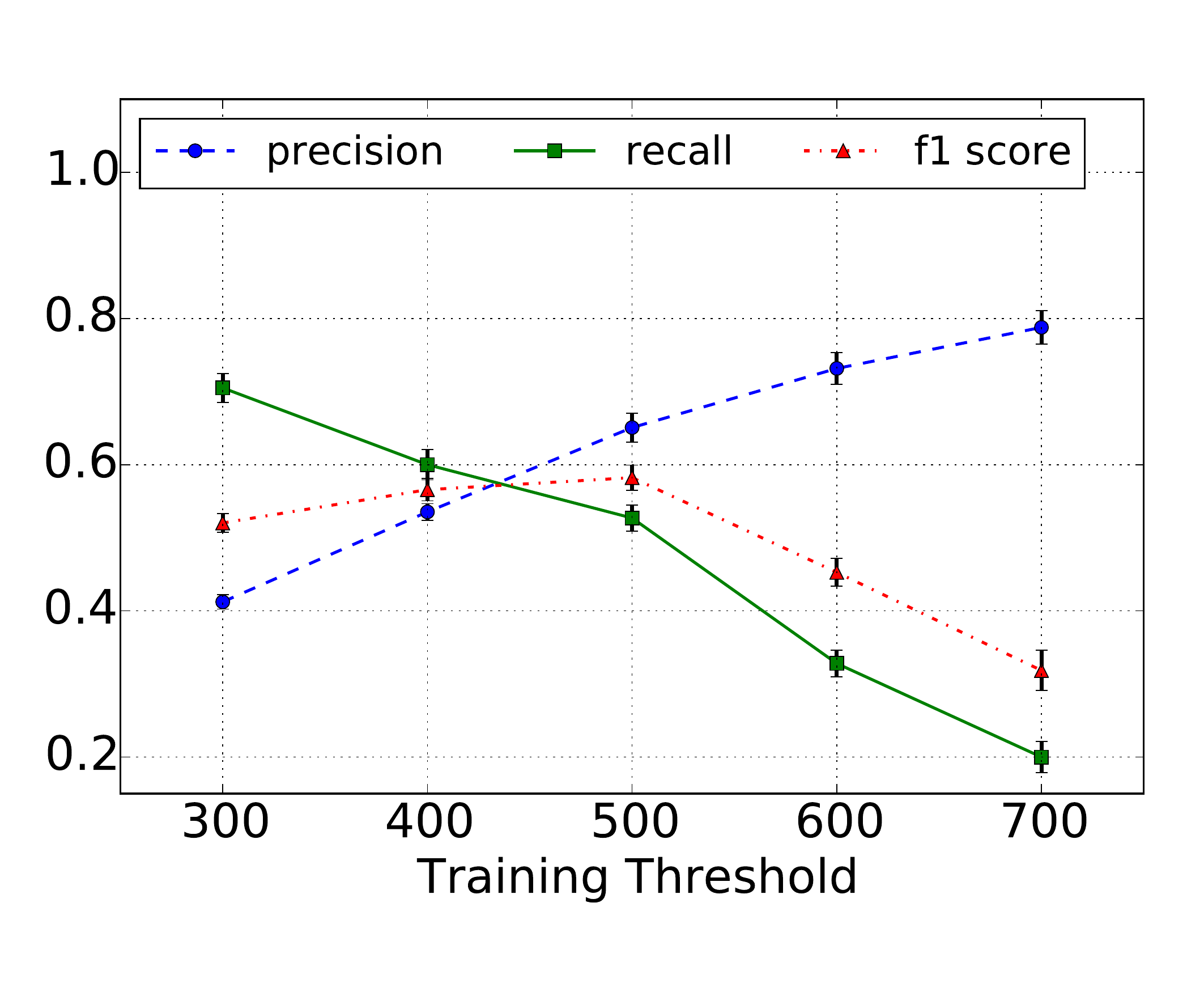}
		\caption{Results for features in $A_m$ with different $TH_{tr}$.}
		\label{fig:to_Size}
		
	\end{subfigure}
	\hfill
	\begin{subfigure}[b]{0.235\textwidth}
		\includegraphics[width=\textwidth]{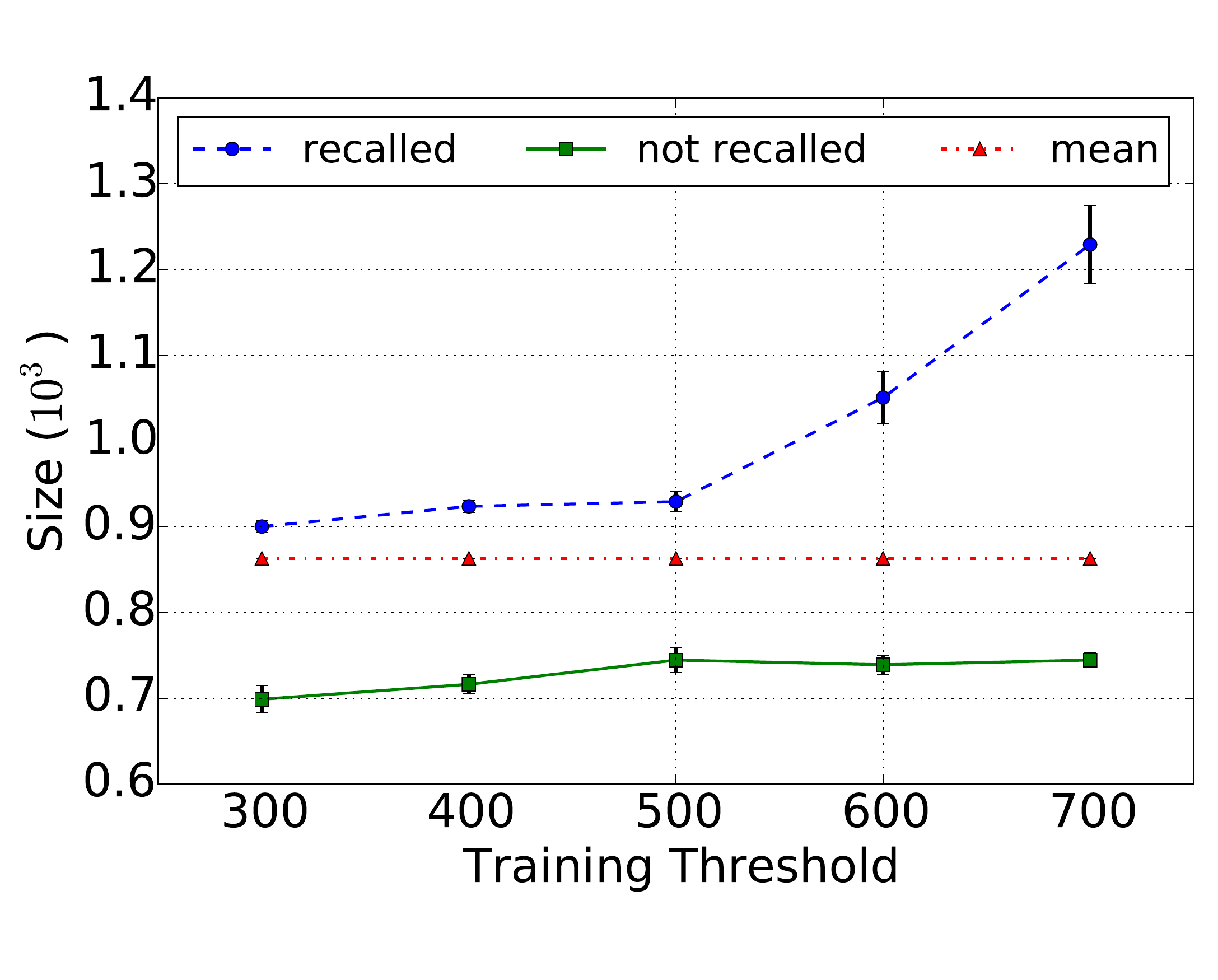}
		\caption{Average final size of viral cascades (recalled, mean and not recalled)}
		\label{fig:tr_th_Size}
		
	\end{subfigure}
	\begin{subfigure}[b]{0.235\textwidth}
		\includegraphics[width=\textwidth]{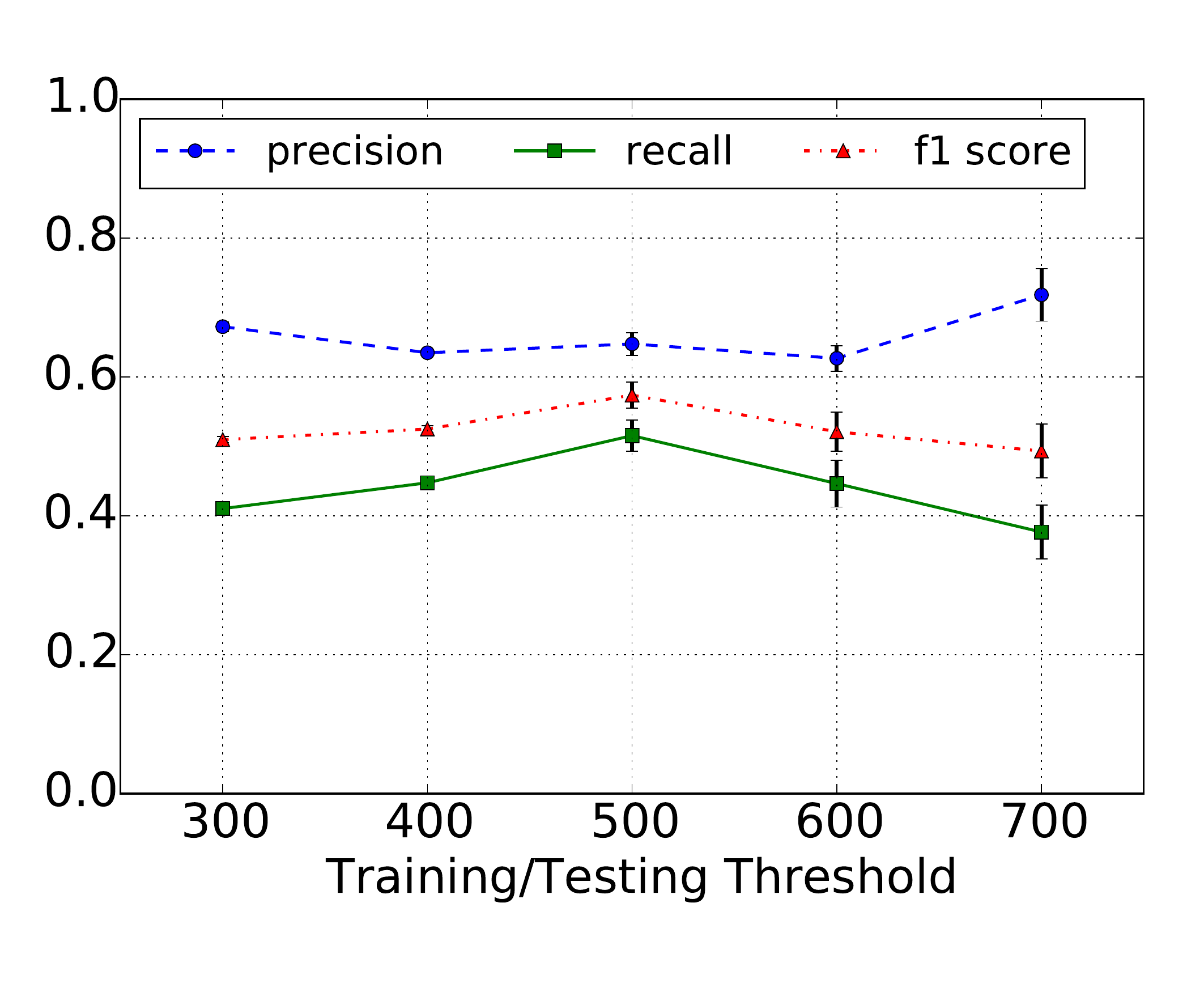}
		\caption{Results for features in group $ A_m $ when $TH_{tr}$ and $TH_{ts}$ change.}
		\label{fig:SIZE_A}
		
	\end{subfigure}
	\hfill
	\begin{subfigure}[b]{0.235\textwidth}
		\includegraphics[width=\textwidth]{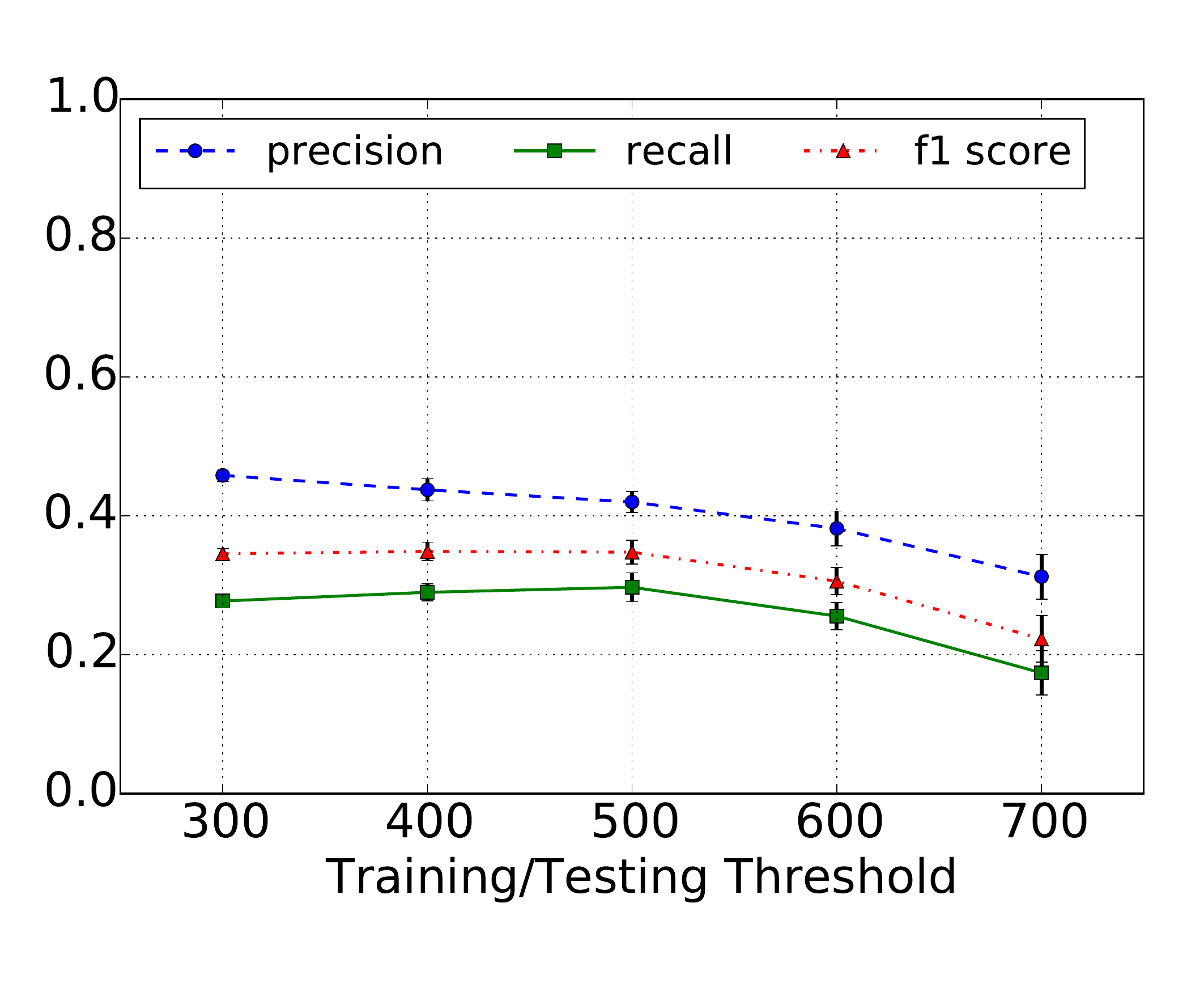}
		\caption{Results for features in group $ B_m $  when $TH_{tr}$ and $TH_{ts}$ change.}
		\label{fig:SIZE_B}
		
	\end{subfigure}
	\caption{Prediction results for $ A_m $ when $TH_{tr}$ and $TH_{ts}$ change. Error bars represent one standard deviation.}
	\label{tngSizeRes}
	
\end{figure}

\noindent\textbf{Size-based prediction.}  We studied cascades of size 50 that reached 500 for this task.  There are 13,285 cascades that can reach the size $ m = 50 $ while 208 out of them reached the size of 500.  Maintaining the threshold $ TH = 500 $, Fig.~\ref{fig:pred_Size} shows random forest classifier trained with features in group $ A_m $ can outperform the other groups.  The trade-off between precision and recall can be achieved by changing the training threshold $TH_{tr}$ while maintaining the testing threshold (Fig.~\ref{fig:to_Size}). We also note that the average final size of viral cascades recalled by the classifier increases with the training threshold (Fig.~\ref{fig:tr_th_Size}).  With threshold $ TH = \left\{300,400,500,600,700\right\} $ on both training and testing samples, the features of group $A_m$ consistently outperform those previously introduced ($B_m$) (Fig.~\ref{fig:SIZE_A}, Fig.~\ref{fig:SIZE_B}).

\noindent\textbf{Feature investigation.}
Here we investigate the importance of each feature in $ A_m $.  With $ TH_{tr} = 500 $ and $ TH_{ts} = 500 $, we trained 100 randomized logistic regressions models - each assigning weights to the features in those sets.  We then categorized the features with weight larger than $ 0.01 $ (on average) into groups such as overlap, gini impurity, etc. Then, we performed classification on the basis of single feature categories or combination of such categories.  The average weights assigned are shown in Table~\ref{tab:fsele}.
As shown by these results, overlaps can make significant contribution to cascade prediction. Intuitively, communication between two sets of nodes is more likely to happen in their shared communities - which is consistent with the results of \cite{ugander2012structural}. This implies that the larger overlap value, the more influence of one set on the other. For example, we can infer that viral cascades tend to have larger $ O(V_{\theta}^m,F_{\theta}^{m,\lambda}) $ value for adopters have larger chance to influence the $\lambda$ frontiers than non-viral cascades.
Moverover, the gini impurity of $\lambda$ non-adopters also shows its importance.
Intuitively, non-viral cascades are easier to be trapped in a relatively small amount of communities. This means even if they could show up in people's timeline with high structural diversity but can not get them infected. 

\begin{table}
	\setlength\extrarowheight{3.5pt}
	\centering
	\setlength\tabcolsep{3pt}
	\begin{minipage}{0.24\textwidth}
		
		\begin{tabular}{|c|c|c|}
			\hline
			Name & Features & Weights \\ \hline
			
			\multirow{3}{*}{Gini} &$I_G(F_{\theta}^{50,\lambda})$ &  0.02 \\ 
			& $I_G(\bar{F}_{\theta}^{50,\lambda})$ & 0.02   \\ 
			& $I_G(\bar{F}_{\theta}^{30,\lambda})$ & \textbf{0.52}   \\ 
			Baseline & $\frac{\sum_{i=1}^{50} t_{\theta}^i}{50} $ & \textbf{1.00} 
			 \\ \hline
		\end{tabular}
	\end{minipage}%
	\hfill
	\begin{minipage}{0.24\textwidth}
		
		\begin{tabular}{|c|c|c|}
		\hline
		 Name & Features & Weights \\ \hline
		 \multirow{5}{*}{Overlap}& $O(V_{\theta}^{30},F_{\theta}^{30,\lambda})$ & \textbf{0.50} \\ 
		 & $O(V_{\theta}^{30},\bar{F}_{\theta}^{30,\lambda})$ & 0.04 \\ 
		 & $O(F_{\theta}^{30,\lambda},\bar{F}_{\theta}^{30,\lambda})$ & 0.23\\ 
		 & $O(V_{\theta}^{50},F_{\theta}^{50,\lambda})$ & \textbf{0.50} \\ 
		 & $O(F_{\theta}^{50,\lambda},\bar{F}_{\theta}^{50,\lambda}) $  & 0.26 \\ \hline	
		\end{tabular}
	\end{minipage}
	\caption{Weights of features assigned by randomized logistic regression models}
	\label{tab:fsele}
\end{table}

\vspace{-5pt}
\section{Acknowledgment}
This work is supported through the AFOSR Young Investigator Program (YIP), grant number FA9550-15-1-0159.

\bibliographystyle{IEEEtran}
\bibliography{ref}

\end{document}